# e-Science perspectives in Venezuela

G. Díaz[1], J. Flórez-López[2], V. Hamar[1], H. Hoeger[3], C. Mendoza[4],
Z. Mendez[1], L. A. Núñez[5], N. Ruiz[1], R. Torrens[1], M. Uzcátegui[1],

[1] *Centro Nacional de Cálculo Científico Universidad de Los Andes (CECALCULA),
Corporación Parque Tecnológico de Mérida, Mérida 5101, Venezuela*
{gilberto,vanessa,nicolas,zulay,torrens,maylett,}@ula.ve

[2] *Departamento de Ingeniería Estructural, Facultad de Ingeniería,
Universidad de Los Andes (ULA), Mérida 5101, Venezuela*
iflorez@ula.ve

[3] *Centro de Simulación y Modelado (CESIMO), Facultad de Ingeniería,
Universidad de Los Andes, Mérida 5101, Venezuela, and
Centro Nacional de Cálculo Científico Universidad de Los Andes (CECALCULA),
Corporación Parque Tecnológico de Mérida, Mérida 5101, Venezuela*
hhoeger@ula.ve

[4] *Centro de Física, Instituto Venezolano de Investigaciones Científicas (IVIC),
PO Box 21827, Caracas, 1020A, Venezuela, and
Centro Nacional de Cálculo Científico Universidad de Los Andes (CECALCULA),
Corporación Parque Tecnológico de Mérida, Mérida 5101, Venezuela*
claudio@ivic.ve

[5] *Centro de Física Fundamental, Departamento de Física, Facultad de Ciencias,
Universidad de Los Andes, Mérida 5101, Venezuela, and
Centro Nacional de Cálculo Científico Universidad de Los Andes (CECALCULA),
Corporación Parque Tecnológico de Mérida, Mérida 5101, Venezuela*
nunez@ula.ve

## Abstract

We describe the e-Science strategy in Venezuela, in particular initiatives by the Centro Nacional de Cálculo Científico Universidad de Los Andes (CECALCULA), Mérida, the Universidad de Los Andes (ULA), Mérida, and the Instituto Venezolano de Investigaciones Científicas (IVIC), Caracas. We present the plans for the Venezuelan Academic Grid and the current status of Grid ULA supported by Internet2. We show different web-based scientific applications that are being developed in quantum chemistry, atomic physics, structural damage analysis, biomedicine and bioclimate within the framework of the E-Infrastructure shared between Europe and Latin America (EELA).

# 1. Introduction

The National Center for Scientific Computing of Los Andes University (*Centro Nacional de Cálculo Científico Universidad de Los Ande*s - CECALCULA) was founded in 1997 as a joint effort between Los Andes University (ULA), the National Scientific and Technological Research Council (*Fondo Nacional de Investigaciones Científicas y Tecnológicas* - FONACIT) and the Mérida Technology Park Corporation (*Corporación Parque Tecnológico de Mérida* - CPTM), with the close cooperation of IBM and Silicon Graphics for the transfer of technology and experimentation in Computational Sciences and Engineering. During this decade, our Center has provided services and staff training in advanced computing. CECALCULA is in practice the first of a national network of decentralized centers of aggregate value of the National Academic Network of Research Centers and National Universities (*Red Académica Nacional de Centros de Investigación y Universidades Nacionales* - REACCIUN) which concentrates efforts and resources (human and infrastructure) to promote the rational use of expensive computer platforms and software under international computational standards. Perhaps the most important feature is that we have created a space for learning and generating a self-organized experience around a high-value technological service. The accumulated experience in the operation of an aggregate value service is only comparable with the management, evaluation and the follow-up of technological projects. The possibility to export and replicate this experience to other institutions is perhaps the greatest challenge to come.

In the past ten years CECALCULA has organized a series of more than a dozen national and regional (Caribbean Basin and Andean countries) workshops and schools aimed at high-level researchers and professionals. It has been a sustained effort to transfer experiences coming from other latitudes to our region. But one of the most important issues was to join the e-Infrastructure shared between Europe and Latin America (EELA) project. Before EELA started, we held the First Latin-American Grid Workshop in Mérida, Venezuela, in November 2004. This event drew a lot of attention and was a first effort to promote grid technology among our countries. EELA has had a great impact in Latin America when it comes to grid technology infrastructure and knowledge. Before EELA, there was almost nobody working locally on grids. Venezuela, together with Brazil and Mexico, participated in an attempt to establish the Latin American and Caribbean Grid (GridLAC: Grid Latino Americano y del Caribe) sponsored by Sun Microsystems, but it never took off. The topic was hot, and when grids were mentioned in workshops and conferences, people got excited and willing to know more about it. However, it was only after EELA was launched that grid operational knowledge really disseminated in the region, not only by means of the many workshops, tutorials and conferences but by working on a real operational grid infrastructure. Now people have grids, know about them and, most important, they are starting to use them in their projects. There is still a lot of work to be done since many more people, institutions and countries are willing to join, and we have to find ways to add more resources to these grids and make them sustainable.

This paper is concerned with what has been so far done regarding grid institutional initiatives based on advanced networking facilities and our plans for the near future. We describe different web-based scientific applications that we are working on in quantum chemistry, atomic physics, structural damage analysis, biomedicine and bioclimate, developed mainly within the framework of EELA.

In Section 2 we discuss our academic grid initiatives both nationally and institutionally. Section 3 describes some of the applications we have been involved with. The regional impact, mainly in Colombia's academic community, is considered in Section 4. Finally, in Section 5 we sketch the conclusions of our experience in collaborating within the EELA project.

## 2. Venezuelan Academic Grid and Grid ULA

### 2.1. Venezuelan Academic Grid

The software, structure and expertise collected in the EELA project, in addition to the various outreach and training activities held in Venezuela and in the region sponsored by EELA, act as a basis and have seeded the development of the Venezuelan academic grid.

The aims of this grid are multiple. Because of the very nature of a grid, namely the sharing of resources, it is expected to lead to a greater degree of cooperation between the participants in addition to the sharing of resources such as clusters.

The grid is being developed in several phases. Initially only a few institutions are being integrated, and after teething problems are sorted out, other institutions will be added.

2.1.1. Phase One

In the first phase the integration of the local grids of the Universidad Simón Bolívar (USB) and the Universidad de Los Andes (ULA) was achieved, with the structure shown in Figure 1.

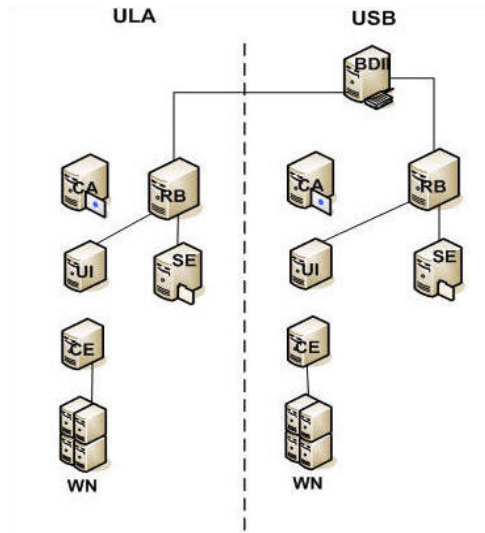

**Figure 1. Grid USB-ULA.**

2.1.2. The Certification Authority

For purposes of security and authentication in grid environments, a certification authority (CA) is essential. In Figure 1, each local grid has its own CA. These CAs are strictly local and their certificates have no validity outside this context. For certificates issued by a CA to be recognized internationally, a CA in Latin America must pass a rigorous scrutiny by The Americas Grid Policy Management Authority (TAGPMA). This process is lengthy and thorough, and only when all the requirements are met, the CA is certified. Currently, ULA has gone through several stages of meetings and revisions, and its CA is expected to be approved soon. As a member of EELA, ULA currently uses certificates issued by other CAs of the consortium that are recognized internationally. Once ULA's CA is approved, it may issue certificates that will be recognized by all sites that integrate EELA and other services that require certificates at an international level.

2.1.3. Phase Two

The second phase was to integrate the site of the Venezuelan Institute for Scientific Research (IVIC) and a computing element (a cluster) of the Universidad Central de Venezuela (UCV) to the grid USB-ULA. We want to emphasize here that to use a grid it is not necessary to have a site. Grid access is through a user interface (UI). An institution that at any given time does not have the resources or does not have the technical capability to install a site can access a grid by installing only a UI or through a UI of another institution.

2.1.4. Phase Three

This phase will integrate other academic institutions to the Venezuelan grid. Figure 2 shows a possible integration scheme and just lists, by way of example and graphics limitations, some institutions.

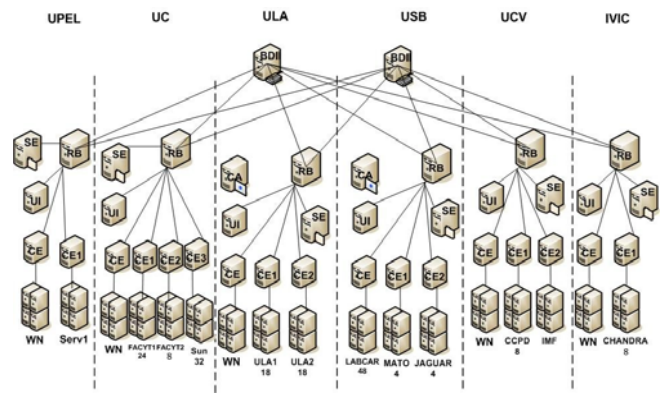

**Figure 2: Venezuelan Academic Grid**

Regarding the final structure that the grid will take, decisions will be taken as institutions are progressively added. Those decisions may include issues such as the following: an institution will either setup a site, only a IU, or access the grid through another institution; which services will be implemented; what will the structure of the grid information system be; how to enforce security and policies; which resources and applications will be provided. All this must go hand in hand with plans for dissemination and training.

**2.2. Grid ULA**

The EELA grid is in itself a grid federation. In ULA's case, the same local infrastructure that supports EELA also backups the Venezuelan academic grid and Grid ULA. Services are shared as well as most of the hardware (computing and storage). User management is carried out in terms of virtual organizations (VO's). Each VO is provided with a set of access rights and each user has to belong to at least one VO to be able to use the grid resources. The user inherits the VO rights.

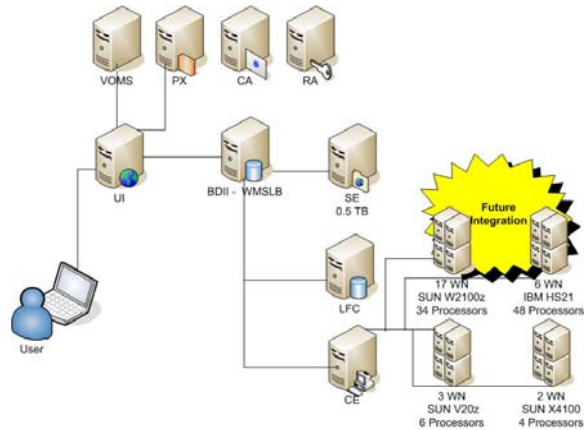

**Figure 3: Current and future stage of Grid ULA**

Besides the hardware on which the grid services are installed, the existing computing facilities consist of 3 Sun V20z (2 processors, 4 GB Ram, 160 GB DD) and 2 Sun X4100 (2 processors, 8 GB Ram, 160 GB DD). 17 Sun W2100z /(2GB Ram and 73 GB DD) should be available in a couple of weeks, and 6 IBM HS21 (2 quad core processors, 16 GB Ram with 146 GB DD) by January 2008. Regarding mass storage, 0.5 TB are provided and 4 TB will be added by January 2008 (see Figure 3).

The generic applications that run on the grid include Siesta, Gaussian 03, Mpi, Scilab, GNU compilers, Abaqus, R, Opendx, SLATEC libraries and PovRay. Besides these applications there are some specific ones which will be mentioned in Section 3.

In order for this significant amount of resources to have a substantial impact on our research communities, the latter need to be educated on how to access and use them. So far training sessions have been held and others are scheduled.

## 3. Some Ongoing Applications

In this section we list some experiences in developing e-Science environments that we have been working on at CECALCULA, most of which have been described elsewhere [1].

**3.1. e-Engineering**

One of the first developments has been the Damage Portal which is a web-based finite element working environment for structural analysis [2]. It allows the user to numerically simulate cracking processes and collapses of reinforced concrete structures subjected to overloads by seismic events. This system consists of a set of Java (working environment) and FORTRAN (generator engine) modules. These modules provide the environment for building the input structure and evaluating it. The simulator is a dynamic, nonlinear finite element program whose physical model is based on a new theory referred to as Lumped Damage Mechanics [3]. The simulator computes and quantifies the density and location of concrete cracking and reinforcement, yielding a set of state variables. In particular, the concrete cracking density is described by a damage variable that can take values between zero (no damage) and one (complete concrete destruction). The Portal has been successfully employed to evaluate the structural risk of educational buildings [4, 5].

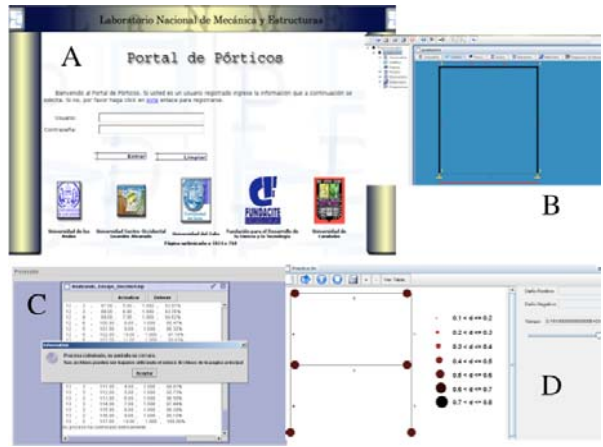

**Figure 4. Damage Portal. Plate A: The Damage Portal Homepage http://portaldeporticos.ula.ve/. Plate B: Pre-processor. Plate C: Monitoring of an analysis through the portal. Plate D: Graphic post-processor**

### 3.2. e-Biomedicine

One of the most common tools in the analysis of nucleotide and protein alignment is BLAST (from its English acronym Basic Local Alignment Search Tool). This computational tool finds similar regions among sequences, compares them with existing sequences in genetic databases and statistically estimates the degree of similarity between them. This homology searching process is computationally demanding, not so much by the search for a particular sequence, but rather because there are usually hundreds of sequences that are explored simultaneously.

### 3.3. Blast2EELA

This is a Biomedical Portal accessed from the URL http://www.cecalc.ula.ve/blast. The bioinformatics community mostly uses local facilities or public servers like NCBI or gPS@, but these environments tend to be inefficient given the limited number of simultaneous searches that can be performed. Additionally databases are updated frequently so it is important to re-check previous results. Within the EELA project, and as one of the biomedical applications [6], the Blast2EELA portal has been developed in cooperation with the Universidad Politécnica de Valencia. This portal is based on the implementation of the parallel BLAST code mpiBLAST [7], and allows simultaneous multiple sequences to be sent for their analysis.

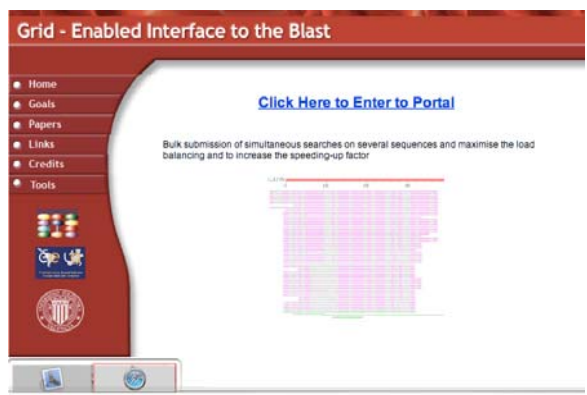

**Figure 5: Blast2EELA Biomedical portal <http://www.cecalc.ula.ve/blast>**

## 3.4. e-Environment

Another initiative to support e-research is the Merida Network of Bioclimatic Stations (REDBC). The access to data sets produced by environmental agencies and institutions in Venezuela is difficult for several reasons: availability of data only on paper, high costs for data acquisition, incomplete or discontinued data and so on. Additionally there re problems related to data capture procedures and data reliability.

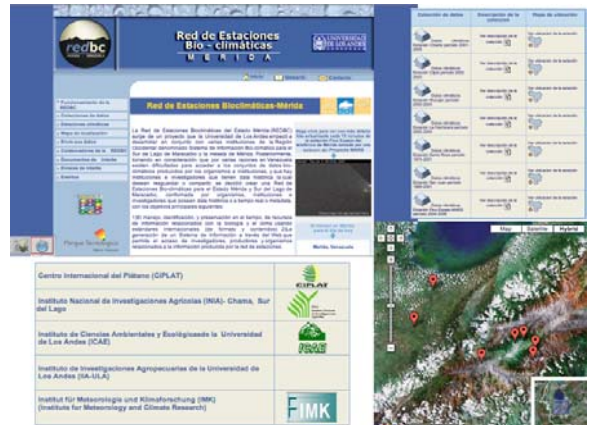

**Figure 6: Merida Network of Bioclimatic Stations (<http://www.cecalc.ula.ve/redbc>). The main portal is shown, the location of the meteorological stations, the data collections and the organizations that collaborate with the network.**

For these reasons we developed the Bioclimatic Network initiative. It is a pilot project which incorporates the capture, processing and dissemination of climatic, environmental and ecological information by implementing an information system and communication mechanisms. Currently, raw data are documented, sorted and published on the network for six (6) weather stations: Chama, Mucujún, La Hechicera, Ciplat, Mucujún, Santa Rosa and San Juan. Additionally didactic information is provided on the generated data and metadata management for the network partners (see Figure 6). This will guarantee data custody and distribution, ensuring retention for long periods of time. Procedures and tools are created to facilitate access to the data generated by any station in the network or by scientists who need them. The network of meteorological stations uses international standards (format and content) to organize and store the data (to allow exchange with other systems). It is designed as a proof of concept and cooperative project, where institutions or individuals with data can document and send them to our center for their preservation. This voluntary approach has been successful since it began with 3 stations; 2 in the southern part of Lake Maracaibo and one on the Merida plateau. Now there are the six stations mentioned above, most of them on the plateau.

## 3.5. From databases to "dataspaces"

The Opacity/Iron Projects [8] have always been concerned with aspects related to data storage, accessibility and dissemination, which, with the advent of the Internet and the web, have reached a high degree of versatility. For instance, in the implementation of an online server of astrophysical opacities, referred to as the OPserver [9], we have considered three user modes (see Figure 7) which illustrate the different ways in which atomic data are required in user applications. In Mode A, the user downloads the complete database locally for intensive access by a modeling code. In Mode B data are also recurrently fetched by the application but this time from a remote central facility through a network subroutine library. This option is practical for a distributed grid environment where the regular network transfer of the complete database, such as in Mode A, would impact performance. Mode C is aimed at the casual user that queries the remote database and downloads data files through a web page.

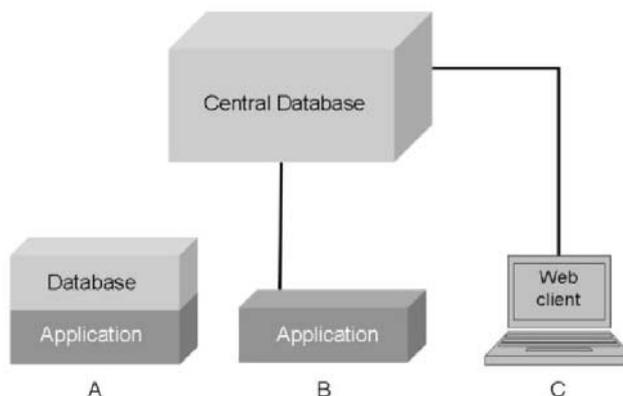

**Figure 7: OPserver user modes. (A) The database is downloaded locally and interfaced with the user's application. (B) The user's application accesses the remote database. (C) Database is accessed through a web page.**

However, these data access modes do not suffice the envisioned e-Science requirements. In future scenarios, most of the data sets will be housed in distributed repositories that do not necessarily comply with a standard data model or the specific structure of a DBMS. Administrative coordination and semantic integration of such data sets can be very low. One must then begin to think in terms of "dataspaces" and Data Space Support Platforms (DSSP) [10] rather than DBMSs. A DSSP must then handle different data formats, interfaces and DBMSs, and allow searches and queries even when not having complete control of the data sets; i.e., a query result may only be approximate. An important point is to develop comprehensive metadata that can be readily picked up by the new generation search engines (e.g., Google), and to implement a custom markup language (XML) [11] for the exchange of atomic and molecular data. An "Atomic and Molecular Data Markup Language"(AMDML) is currently being developed by an international group led by Yuri Ralchenko (NIST, USA) which will give atomic data sets a flexible yet standardized semi-structure and promote their cataloguing in Google searches.

## 4. Regional Cooperation

One of the most important byproducts of the EELA project has been the promotion of regional advanced networking organizations. Even in those countries that initially were not associated with EELA, the project has leveraged the generation of such organizations and the dissemination of computer skills. This is the case of Colombia, where several universities and regional high speed networking are now working together with CECALCULA in order to build the Grid Colombia.

Since 2002 CECALCULA has organized several workshops at the Universidad Industrial de Santander, and more recently with the support of the EELA, we co-organized the fourth EELA workshop and the 9th EELA tutorial (March 2007) at the Universidad de Los Andes in Bogota (UniAndes). We are assisting them, together with the Universidade Federal do Rio de Janeiro in Brazil, in installing a complete EELA grid site. CECALCULA also co-organized the first edition of the Latin American Conference on High Performance Computing in Santa Marta,

Colombia, as a forum to cultivate the work that is being done in Latin America in the field. Next editions are already scheduled for Mexico and Venezuela.

## 5. Conclusions

In this work we have shown our plans to establish a Venezuelan Academic Grid in order to promote the sharing of resources, knowledge and foster the cooperation among Venezuelan universities and research institutions. On the other hand, we have also presented the current stage of Grid ULA and near-future enhancements. Both initiatives are highly motivated by the know-how gained from the EELA project.

Some developments of different applications that are becoming available through user-friendly portals and that require the computing power and/or the distribution that a grid offers are also presented. Lastly, we shortly describe how CECALCULA is helping to promote and provide training in the region in this kind of advanced computing technology.

## References

[1] J.L. Chaves, G. Díaz, V. Hamar, R. Isea, F. Rojas, N. Ruíz, R. Torrens, M. Uzcátegui, J. Flórez-López, H. Hoeger, L. Núñez, and C. Mendoza. e-science initiatives in venezuela. In Proceedings of the *Spanish Conference on e-Science Grid Computing*, J. Casado, R. Mayo y R. Muñoz (editors), pp 45 - 52 CIEMAT Madrid Spain, 2007.

[2] M. E. Marante, L. Suárez, A. Quero, J. Redondo, B. Vera, M. Uzcategui, S. Delgado, L. R. León, L. Núñez, and J. Flórez-López. Portal of damage: a web-based finite element program for the analysis of framed structures subjected to overloads. *Advances in Engineering Software*, 36:346 –358, 2005.

[3] A. Cipollina, A. López-Inojosa, and J. Flórez-López, A simplified damage mechanics approach to nonlinear analysis of frames. *Comp. & Struct., Struct.*, 54:1113 – 1126, 1995.

[4] A. Moreno. Influencia del factor de reducción de respuesta en el daño estructural de pórticos de concreto armado sometido a solicitaciones sísmicas. Master's thesis, Facultad de Ingeniería, Universidad del Zulia, 2005.

[5] M. Torres, P. Gonzales, and J. Mujica. Evaluación de la vulnerabilidad estructural de edificaciones esenciales: caso hospital rotario. Trabajo especial de grado, Universidad Centro Occidental Lisandro Alvarado, Barquisimeto Venezuela, 2007.

[6] M. Cárdenas, V. Hernández, R. Mayo, I. Blanquer, J. Pérez-Griffo, R. Isea, L. Núñez, H. R. Mora, and M. Fernández. Biomedical aplications in eela., *Stud Health Technol Inform*, **120**:397–400, 2006.

[7] Y. J. Kim, A. Boyd, B. D. Athey, and J. M. Patel. miblast: scalable evaluation of a batch of nucleotide sequence queries with blast. *Nucleic Acids Res.*, **33**:4335–4344, 2005.

[8] http://cdsweb.u-strasbg.fr/topbase/op.html and http://www.usm.uni-muenchen.de/people/ip/iron-project.html

[9] C. Mendoza, et al., MNRAS, in press (preprint astro-ph/0704.1583), 2007.

[10] M. Franklin, A. Halevy A, D. Maier, SIGMOD Record, **24**, 27, 2005.

[11] J. Freire, M. Benedikt, Comp. Sc. Eng., **6**, 12, 2004.